\documentclass[aps,twocolumn,superscriptaddress]{revtex4}
\usepackage{graphicx}
\usepackage{amsmath}
\usepackage{xcolor}
\begin{document}
\title{Spatial dissipative solitons in graphene-based active random metamaterials}
\author{Ashis Paul} 
\email{apaul@phy.iitkgp.ac.in}
\affiliation{Department of Physics, Indian Institute of Technology Kharagpur, West Bengal 721302, India}
\author{Andrea Marini} 
\email{andrea.marini@univaq.it}
\affiliation{Department of Physical and Chemical Sciences, University of L'Aquila, Via Vetoio, 67100 L'Aquila, Italy}
\author{Samudra Roy}
\email{samudra.roy@phy.iitkgp.ac.in}
\affiliation{Department of Physics, Indian Institute of Technology Kharagpur, West Bengal 721302, India}

\begin{abstract}
We investigate dissipative nonlinear dynamics in graphene-based active metamaterials composed of randomly dispersed graphene 
nano-flakes embedded within an externally pumped gain medium. We observe that graphene saturable nonlinearity produces a sub-critical
bifurcation of nonlinear modes, enabling self-organization of the emitted radiation into several dissipative soliton structures with 
distinct topological charges. We systematically investigate the existence domains of such nonlinear waves and their spatio-temporal dynamics, 
finding that soliton vortices are unstable, thus enabling self-organization into single dissipative structures with vanishing topological charge, 
independently of the shape of the graphene nano-flakes. Our results shed light on self-organization of coherent radiation structures in disordered 
systems and are relevant for future cavity-free lasers and amplifier designs.
\end{abstract}
\maketitle

\section{Introduction} 

Dissipative solitons (DSs) are self-organized localized structures arising in diverse dissipative physical and biological systems \cite{AkhmedievBook},
playing a central role in mode-locked lasers \cite{GreluNatPhot2012,PengSciAdv2019}, photonic molecules \cite{HelgasonNatPhot2021}, time-delay feedback systems \cite{YanchukPRL2019}, Bose-Einstein condensates of exciton polaritons \cite{Ostrovskaya2012} and cold atoms \cite{BrazhniPRL2009,TesioOptExpr2013}. Furthermore, DSs offer an appealing platform for cavity-free stimulated emission of radiation, as in particular in optical amplifiers \cite{UltanirPRL2003}. In this context, innovative cavity-free radiation sources can be achieved also in colloidal solutions embedding randomly arranged scatterers and an optically active medium \cite{WiersmaNatPhys2008,Gottardo2008,Wiersma2013} - the so-called random lasers (RLs). However, cavity-free stimulated emission of radiation in RLs implies poor quality of the output beam, which inherently lacks reproducibility and tunability. 

Recently, the advent of nanophotonic materials has enabled plasmon stimulated emission in metallic nanoparticles \cite{Bergman2003,Noginov2009,Stockman2010} and waveguides embedding gaining media 
\cite{Noginov2008,Marini2009,Bolger2010,DeLeon2010}. Metamaterials (MMs) constitute a promising platform for planar sources of coherent radiation \cite{Zheludev2008} and for cavity-free lasers operating in the stopped-light regime \cite{Hess2012,Marini2016}. Furthermore, MMs embedding subwavelength randomly arranged nanostructures in an amplifying medium are promising for achieving self-organized stimulated emission \cite{MariniPRL2016}, enabling to control the properties of the output beam thanks to stable DSs. This requires the engineering and exploitation of the MM effective nonlinear optical response, which is greatly facilitated by graphene \cite{Bonaccorso2010,Bao2012,Javier2014} because it provides saturated absorption at low peak intensities $\simeq$ MW$/$cm$^2$ \cite{Bao2009,Xing2010,MariniPRB2017} over a broad spectrum spanning the optical and infrared frequency ranges. This peculiar property of graphene is enabled by its conical band-structure providing resonant interband absorption at arbitrary radiation frequencies, thus greatly facilitating the engineering of the random MM effective nonlinear response \cite{MariniPRL2016}.

Here, we investigate the dissipative dynamics of nonlinear waves emitted in a realistic MM composed of disordered graphene nano-flakes embedded in externally pumped rhodamine 6G (R6G). Our model, accounting
non-perturbatively for the saturable absorption of graphene nano-flakes and for the saturable gain of R6G, is based on an effective nonlinearly-saturated propagation equation for the field envelope of the seeded radiation. We observe that homogeneous nonlinear waves (HNWs) bifurcate from the trivial background sub-critically and that are bistable in a particular range of the linear gain coefficient. We further investigate HNW modulational instability, calculating the modulation gain spectrum, which suggests that DSs can be excited in the system. In turn, we investigate numerically DS existence, finding DS vortices with topological charges $m=0,1,2$ and determining their existence domains. We finally investigate DS stability over propagation, finding that only the $m=0$ DSs can be stable in a well-determined range of the linear gain coefficient. Our results extend previous $(1+1)$D investigations \cite{MariniPRL2016} shedding light to nonlinear dynamics in $(2+1)$D and indicating that DSs offer a viable platform to manipulate the spatial pattern of the radiation seeded in random active MMs.

\section{Model} 

\begin{figure}[t]
\centering
\begin{center}
\includegraphics[width=0.5\textwidth]{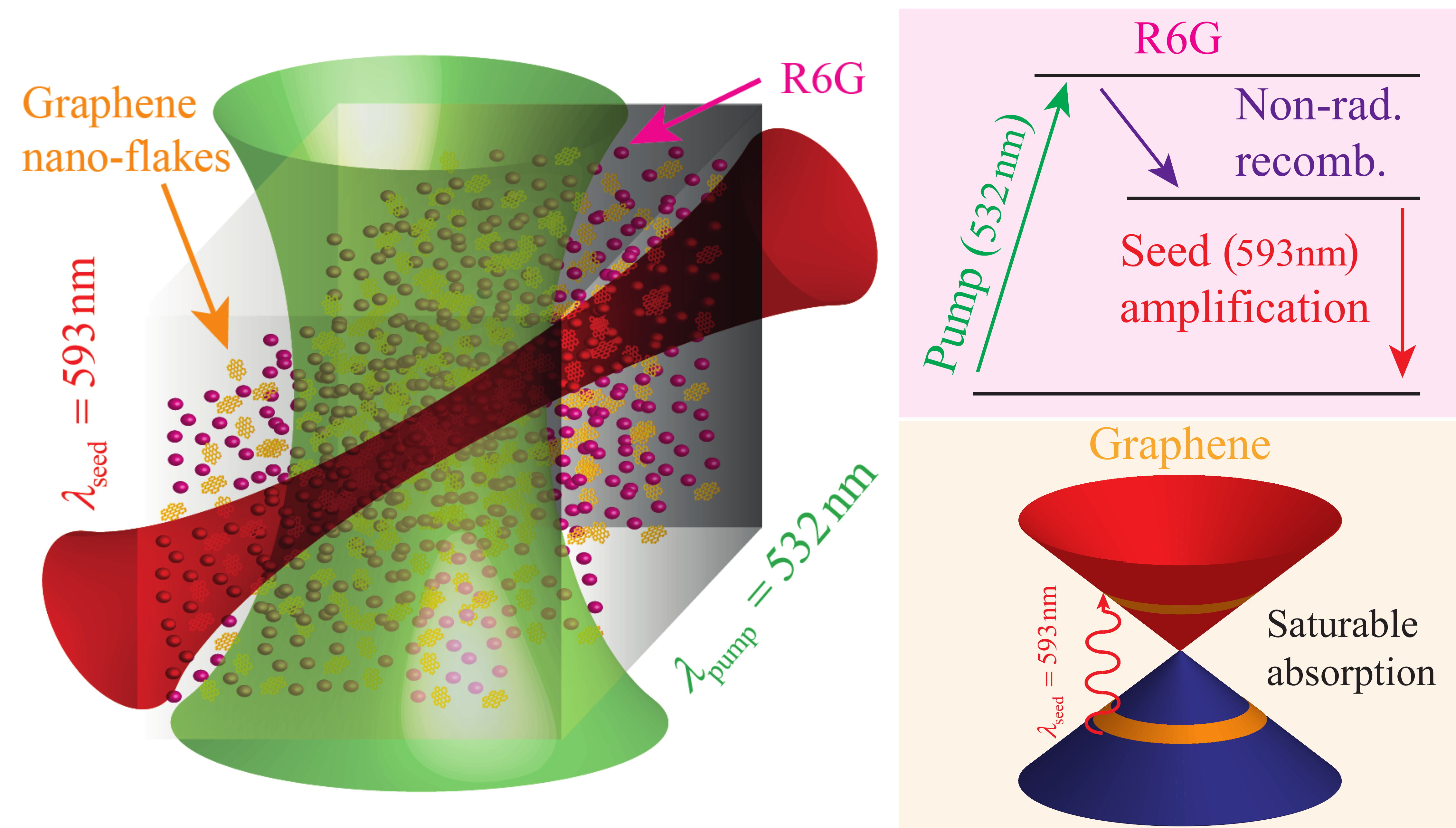}
\caption{Sketch of the considered graphene-based active MM consisting of a mixture of graphene nano-flakes and optically pumped R6G molecules embedded in PMMA.}
\label{Fig1}
\end{center}
\end{figure}

We consider a disordered medium, schematically depicted in Fig. \ref{Fig1}, composed of undoped graphene nano-flakes (with lateral dimensions $< 50$ nm) dispersed in polymethyl methacrylate (PMMA) embedding optically pumped R6G dyes, a typical gaining medium for RLs, see, e.g., Ref. \cite{Leonetti2013}. Population inversion in R6G can be attained by a frequency-doubled Nd-YAG laser-pump beam with wavelength $\lambda_{\rm pump} = 532$ nm, providing peak stimulated emission at $\lambda_{\rm seed} = 593$ nm. In turn, owing to the deep subwavelength dimensions of graphene nano-flakes, radiation seeded in such a system experiences an effective dissipative and nonlinear dielectric response arising from electron dynamics in graphene nano-flakes, PMMA and R6G. 

\begin{figure}[t]
\centering
\begin{center}
\includegraphics[width=0.5\textwidth]{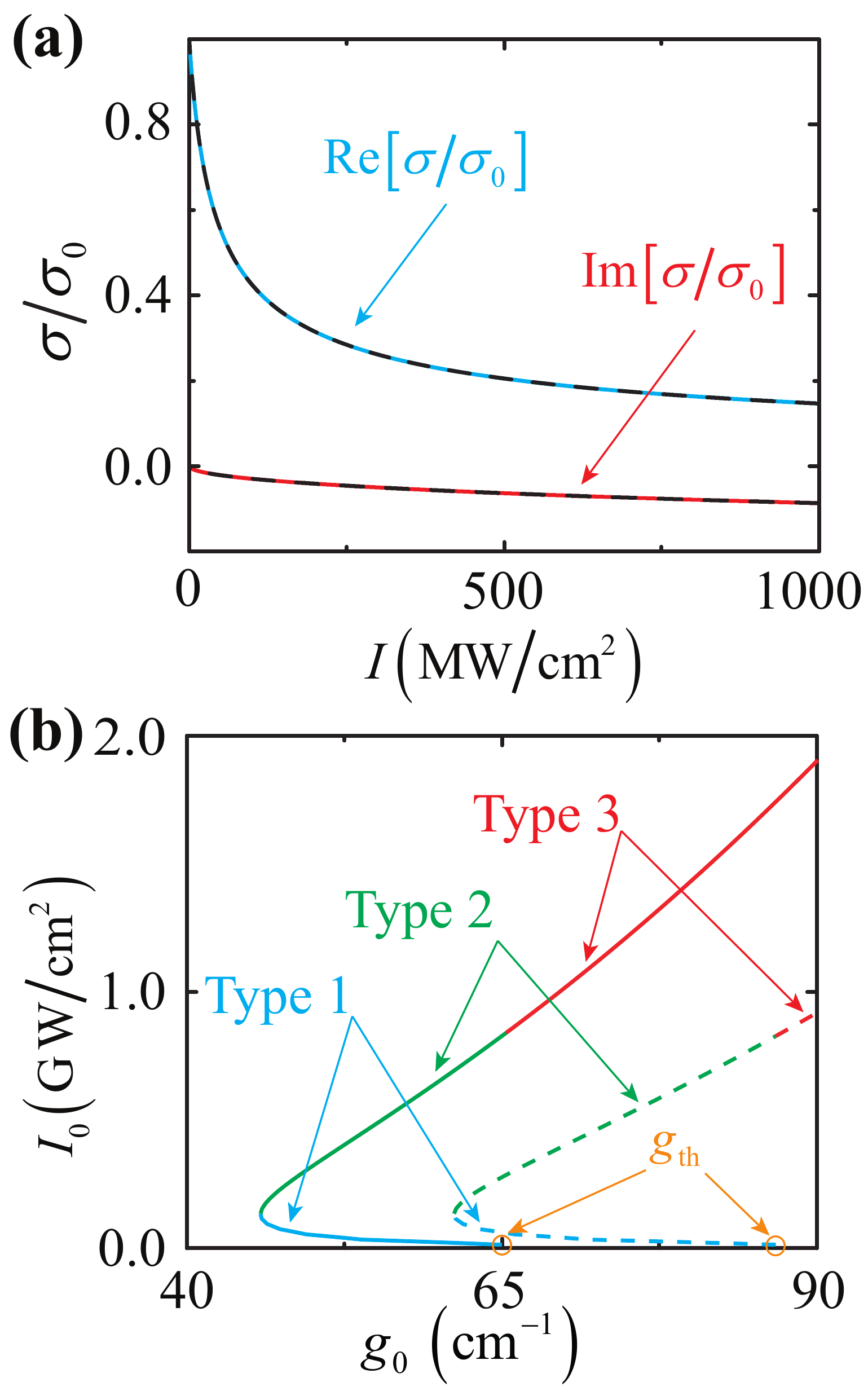}
\caption{{\bf (a)} Dependence of the real (${\rm Re}[\sigma / \sigma_0]$, blue line) and imaginary (${\rm Im}[\sigma / \sigma_0]$, red line) parts of the 
nonlinear conductivity $\sigma$ of undoped graphene nano-flakes (normalized to the linear conductivity $\sigma_0$) over the intensity $I$ of seed radiation 
at $\lambda_{\rm seed} = 593$ nm. Dashed lines indicate fitting with the expression 
$\sigma (I) = \sigma_0 [ \frac{1}{ \sqrt{ 1 + I/I_{\rm S} } } - i \frac{ 1 - e^{ - \eta_1 \sqrt{ I/I_{\rm S} } } }{ \sqrt{1 + \eta_2 (I/I_{\rm S})^{0.4} } }]$.  
{\bf (b)} Existence curves of stationary HNWs, where $I_0 = (1/2)\epsilon_0 \sqrt{\epsilon_{\rm b}} c |A_0|^2$ and $g_0$ is the linear gain coefficient for 
$f = 2.13 \times 10^{-4}$ (full lines) and $f = 2.84 \times 10^{-4}$ (dashed lines). The orange circle dots indicate the background instability threshold 
$g_{\rm th}$ for the two distinct filling fractions considered.}
\label{Fig2}
\end{center}
\end{figure}

PMMA acts as a background dielectric with relative permittivity $\epsilon_{\rm b} \simeq 2.23$. Because of the subwavelength dimensions, the intensity-dependent effective 
dielectric response $\epsilon_{\rm eff}(I)$ (where $I$ is the intensity of the seed beam) of the considered disordered system does not depend on the geometrical details 
of the graphene nano-flakes but only on their filling fraction $f=V_{\rm gr}/V$. Furthermore, because unintentional doping of the graphene nano-flakes ($E_{\rm F} \simeq 0.2$ eV) 
does not affect significantly $\epsilon_{\rm eff}$ in the visible, we assume vanishing doping $E_{\rm F}=0$. We model electron-dynamics in graphene following a previously 
reported non-perturbative approach in thee massless Dirac fermion picture \cite{MariniPRB2017} and calculate numerically the intensity-dependent conductivity $\sigma(I)$, 
which real and imaginary parts are depicted in Fig. \ref{Fig2}a by the full lines. We find good fitting with the analytical expression 
$\sigma (I) = \sigma_0 [ \frac{1}{ \sqrt{ 1 + I/I_{\rm S} } } - i \frac{ 1 - e^{ - \eta_1 \sqrt{ I/I_{\rm S} } } }{ \sqrt{1 + \eta_2 (I/I_{\rm S})^{0.4} } }]$, where 
$\sigma_0 = e^2/4\hbar$, $I_{\rm S} = 137 \hbar \omega_{\rm S}^2 \omega_{\rm seed}^2 / (8\pi v_{\rm F}^2)$, 
$\omega_{\rm S} = 6.16$ rad ps$^{-1}$, $\eta_1 = (\omega_\eta/\omega_{\rm seed})$, $\eta_2 = (\omega_\eta/\omega_{\rm seed})^{0.8}$, $\omega_\eta = 46.20$ rad ps$^{-1}$, $e$ is 
the electron charge, $\omega_{\rm seed} = 2 \pi c/\lambda_{\rm seed}$, $c$ is the speed of light in vacuum, $\hbar$ is the reduced Planck constant and $v_{\rm F}\simeq c/300$ 
is the Fermi velocity of electrons in graphene.

\begin{figure*}[t]
\centering
\begin{center}
\includegraphics[width=\textwidth]{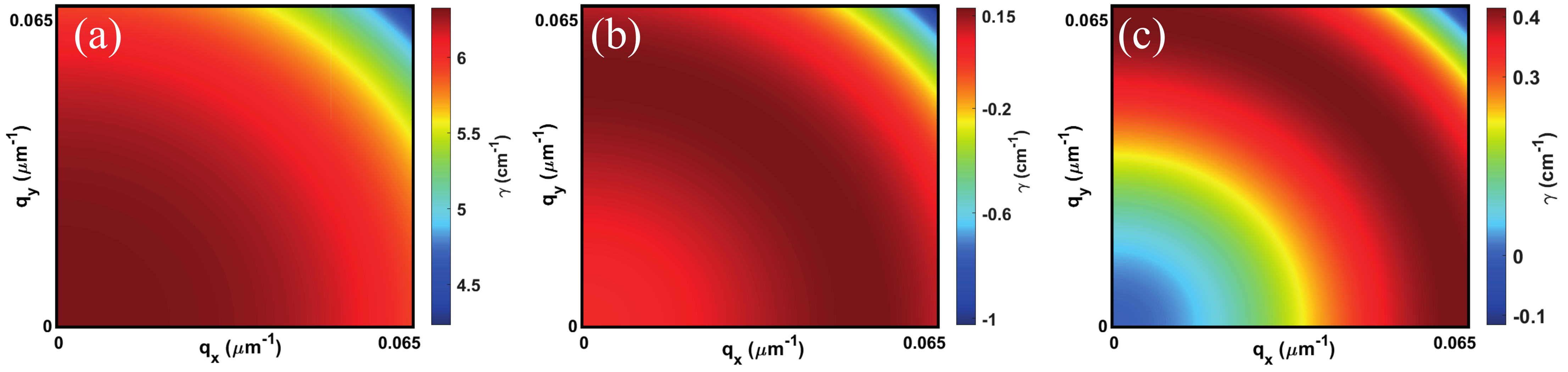}
\caption{Gain spectrum of {\bf (a)} Type I for $g_0 = 60$ cm$^{-1}$, {\bf (b)} Type II for $g_0 = 60$ cm$^{-1}$, and {\bf (c)} Type III for $g_0 = 80$ cm$^{-1}$ HNWs, depicting
the maximum instability growth rate $\gamma$ as a function of $q_x$ and $q_y$ for $f = 2.13 \times 10^{-4}$.}
\label{Fig3}
\end{center}
\end{figure*}

Thus, the induced polarization of graphene nano-flakes with filling fraction $f$ is accounted by the susceptibility $\chi_{\rm G} = (2/3) i f \sigma(I)\lambda_{\rm seed}/(2\pi t_{\rm gr}\epsilon_0 c)$, where $\epsilon_0$ is the vacuum dielectric permittivity, $t_{\rm gr}=0.335$ nm is the graphene effective thickness taken from the 
interlayer distance in graphite, and the factor $2/3$ accounts for averaging over the random orientation of the nano-flakes. Amplification by R6G is modeled by a two-level system,
where stable population inversion is assumed to be produced by the external optical pump operating at $\lambda_{\rm pump} = 532$ nm. Because the seed photon is resonant with the 
two-level transition energy, R6G produces a purely imaginary susceptibility $\chi_{\rm R6G} = - i ( g_0(I_{\rm pump}) / k_{\rm seed}) / [1 + I/I_{\rm S}^{\rm R6G}]$, where 
$k_{\rm seed} = 2\pi / \lambda_{\rm seed}$, $I_{\rm S}^{\rm R6G} \simeq 150$ MW$/$cm$^2$ is the R6G saturation intensity \cite{Nithyaja2011}, and 
$g_0(I_{\rm pump}) \simeq 400$ cm$^{-1}$ is the R6G linear gain coefficient \cite{Noginov2008} that depends over the R6G density and can be tuned by the external pump. 
Thus, the effective response of the complex system depicted in Fig. \ref{Fig1} in the limit of small graphene density is accounted by the complex and nonlinear dielectric 
constant $\epsilon_{\rm eff} (I) \simeq \epsilon_{\rm b} + \chi_{\rm R6G} + \chi_{\rm G}$. 

We model nonlinear propagation of the monochromatic seed beam with angular frequency $\omega_{\rm seed}$ and carrier wave-number 
$\beta_0 = k_{\rm seed} \sqrt{\epsilon_{\rm b}}$ by taking the Ansatz ${\bf E}_{\rm seed}({\bf r},t) = {\rm Re} [ A ({\bf r}_{\bot},z) e^{i\beta_0 z - i\omega_{\rm seed} t} {\bf n} ]$, 
where ${\bf r} = ({\bf r}_{\bot}, ~ z)$ is the position vector, ${\bf n}$ is the seed polarization unit vector, and $A ({\bf r}_{\bot},z)$ is the field envelope. 
In the slowly varying envelope approximation (SVEA) \cite{Leonettii2013}, Maxwell's equations reduce to a generalized Ginzburg-Landau equation \cite{MariniPRL2016} 
for the field envelope 
\begin{equation}
i \partial_z A + \frac{1}{2\beta_0}\nabla^2_{\bot} A + \frac{\beta_0}{2\epsilon_{\rm b}} \left[\epsilon_{\rm eff} \left(|A|^2\right) - \epsilon_{\rm b} \right] A = 0. \label{PropEq}
\end{equation}
The nonlinear propagation above accounts for the effects of diffraction, R6G gain, and graphene saturable absorption on the spatial evolution of the seed beam in the effective
medium approximation.

\section{Nonlinear waves}

\subsection{HNWs}

Extended HNWs are excited by a seed plane wave where the input envelope does not depend over the transverse position $A({\bf r}_{\bot},0) = A_0$.
We calculate the nonlinear dispersion of stationary HNWs by taking the Ansatz $A(z) = A_0 e^{i\delta\beta z}$ in Eq. (\ref{PropEq}) suppressing
the diffraction term $\nabla^2_{\bot} A = 0$ and obtaining the propagation constant correction 
$\delta\beta = (\beta_0/2\epsilon_{\rm b}) \left\{ {\rm Re} \left[\epsilon_{\rm eff} \left(|A_0|^2\right) \right] - \epsilon_{\rm b}\right\}$.
Conversely to Hamiltonian systems where $A_0$ is arbitrary, owing to dissipation (accounted by ${\rm Im}\epsilon_{\rm eff}$) stationary HNWs exist only for specific 
input amplitudes $A_0$ fixed by the nonlinear condition ${\rm Im} [\epsilon_{\rm eff}\left(|A_0|^2\right)] = 0$, which we solve by the Newton-Raphson method 
for several distinct values of the linear gain coefficent $g_0$. Thanks to graphene saturable absorption, we observe a subcritical bifurcation from the background $A_0 = 0$
at a specific gain threshold $g_{\rm th}$ depending over the graphene filling fraction $f$ and leading to bistability of two distinct solutions for $g_0 < g_{\rm th}$, 
which we name type I and II, as illustrated in Fig. \ref{Fig2}b. Conversely, for $g_0 > g_{\rm th}$ we find only one homogenous solution, which we name Type III, 
see Fig. \ref{Fig2}b.

\subsection{Modulational instability of HNWs}

The stability of HNWs against small-amplitude perturbing waves with amplitudes $\delta A_1$, $\delta A_2$ and wave-vector ${\bf q}_{\bot}$ is evaluated by setting
\begin{eqnarray}
A= A_0 + \big[\delta A_1 {\rm e}^{ h z + i {\bf q}_{\bot}\cdot{\bf r}_{\bot}} +  \delta A_2^* {\rm e}^{h^*z - i {\bf q}_{\bot}\cdot{\bf r}_{\bot}}\big] {\rm e}^{i \delta \beta z}, 
\end{eqnarray}
 where $h= \gamma + i \Upsilon $ ,  $\gamma$ and $\Upsilon$ are the instability growth rate and the wave-number shift, respectively, induced by the perturbing waves. Inserting this expression in 
Eq. (\ref{PropEq}), and linearizing with respect to the small-amplitudes $\delta A_1, \delta A_2$, we find
\begin{equation}
\begin{bmatrix}-iF_1+iF_2 & i\delta_2 F' A_0^2 \\ - i\delta_2 F'^{*} A_0^{2 *} & iF_1+iF_2^* \end{bmatrix} \begin{bmatrix} \delta A_1 \\ \delta A_2 \end{bmatrix}= h \begin{bmatrix} \delta A_1 \\ \delta A_2 \end{bmatrix} 
\end{equation}
where, $\delta_2=(2\beta_0)^{-1}$, $F=(\beta_0/2\epsilon_b)\big[\epsilon_{eff} (|A_0|^2)-\epsilon_b\big]$, $F'= \partial_{|A_0|^2} F$, $F_1 = \delta \beta + \delta_2 q_{\perp}^2$ and $F_2=\delta_2 (F' |A_0|^2+F)$.

\begin{figure*}[t]
\centering
\begin{center}
\includegraphics[width=\textwidth]{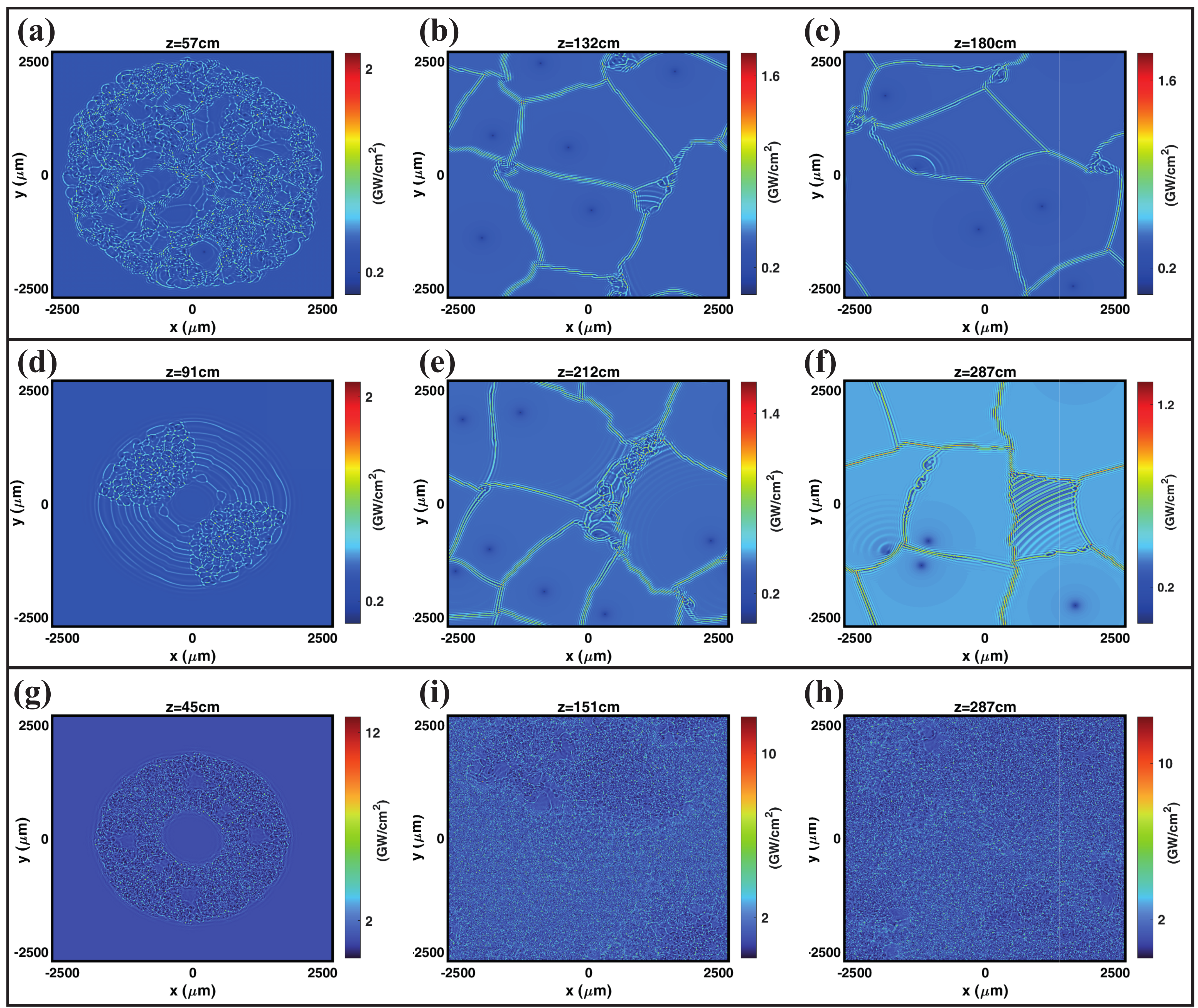}
\caption{Evolution of perturbed HNWs of {\bf (a-c)} Type I ($g_0 = 60$ cm$^{-1}$), {\bf (d-f)} Type II ($g_0 = 60$ cm$^{-1}$), and {\bf (g-i)} Type III 
($g_0 = 80$ cm$^{-1}$) for $f = 2.13 \times 10^{-4}$. The density plots depict the dependence of the intensity profile $I({\bf r}_{\bot})$ over the transverse
position vector ${\bf r}_{\bot} = x\hat{e}_x + y\hat{e}_y$ for several distinct propagation distances $z=\bar{z}$ indicated at the top of every figure.}
\label{Fig4}
\end{center}
\end{figure*}

We calculate numerically the complex eigenvalues $h=\gamma + i \Upsilon$ of such a linear algebraic system of equations for every wave-vector ${\bf q}_{\bot}$. 
Positive/negative growth rates $\gamma$ indicate instability/stability against small-amplitude perturbations. We find that all HNWs are unstable and that, while 
the maximum instability growth rate $\gamma$ is peaked at $q_{\bot} = 0$ for Type I I HNWs, see Fig. \ref{Fig3}a, for Type II and Type III HNWs it is peaked at a specific 
$q_{\bot} = \bar{q}_{\bot}$, see Figs. \ref{Fig3}b,c. This suggests that perturbed HNWs can develop filamentation over propagation. Furthermore, we evaluate 
the stability of the background $A_0 = 0$, finding that it is stable for $g_0<g_{\rm th}$ and unstable for $g_0>g_{\rm th}$. In turn, $g_{\rm th}$ represents 
the linear gain stability threshold of the background. Owing to the above described modulational instability scenario, nonlinear dynamics in subcritical 
and overcritical domains leads to qualitatively different phenomena. In order to gain insight on the evolution of perturbed HNWs, we solve Eq. (\ref{PropEq}) 
in propagation by the split-step discrete Fourier transform complemented with a fourth-order Runge-Kutta algorithm, which results are summarized in 
Fig. \ref{Fig4}, where we depict the intensity profile $I({\bf r}_{\bot},\bar{z})=(1/2)\epsilon_0 \sqrt{\epsilon_b}c|A({\bf r}_{\bot},\bar{z})|^2$ as a function 
of ${\bf r}_{\bot}$ for (a-c) Type I ($g_0 = 60$ cm$^{-1}$), (d-f) Type II ($g_0 = 60$ cm$^{-1}$), and (g-i) Type III ($g_0 = 80$ cm$^{-1}$) perturbed HNWs 
for $f = 2.13 \times 10^{-4}$ at several distinct propagation distances $z=\bar{z}$. Note that for $g_0 = 60$ cm$^{-1}$ (subcritical since $g_0<g_{\rm th}$, 
see Fig. \ref{Fig1}b) both Type I and Type II HNWs develop instabilities over propagation and tend to filament into self-organized patterns and interacting 
domains, while for $g_0 = 80$ cm$^{-1}$ (overcritical since $g_0>g_{\rm th}$, see Fig. \ref{Fig1}b) the dynamics becomes chaotic \cite{MariniPRA2010}.

\subsection{Localized NWs}

Localized, stationary and self-sustaining nonlinear waves, commonly named dissipative solitons (DSs) \cite{Grelu2012}, are calculated by setting the Ansatz 
$A({\bf r}_{\bot},z) = A_0(r_{\bot}) {\rm e}^{i\delta\beta z+im\phi}$ in Eq. (\ref{PropEq}), where $m$ is an arbitrary topological charge and $\phi$ is the 
azimuthal angle between ${\bf r}_{\bot}$ and the $x$-axis, obtaining an ordinary differential equation (ODE) for $A_0(r_{\bot})$, 

\begin{figure}[t]
\centering
\begin{center}
\includegraphics[width=0.5\textwidth]{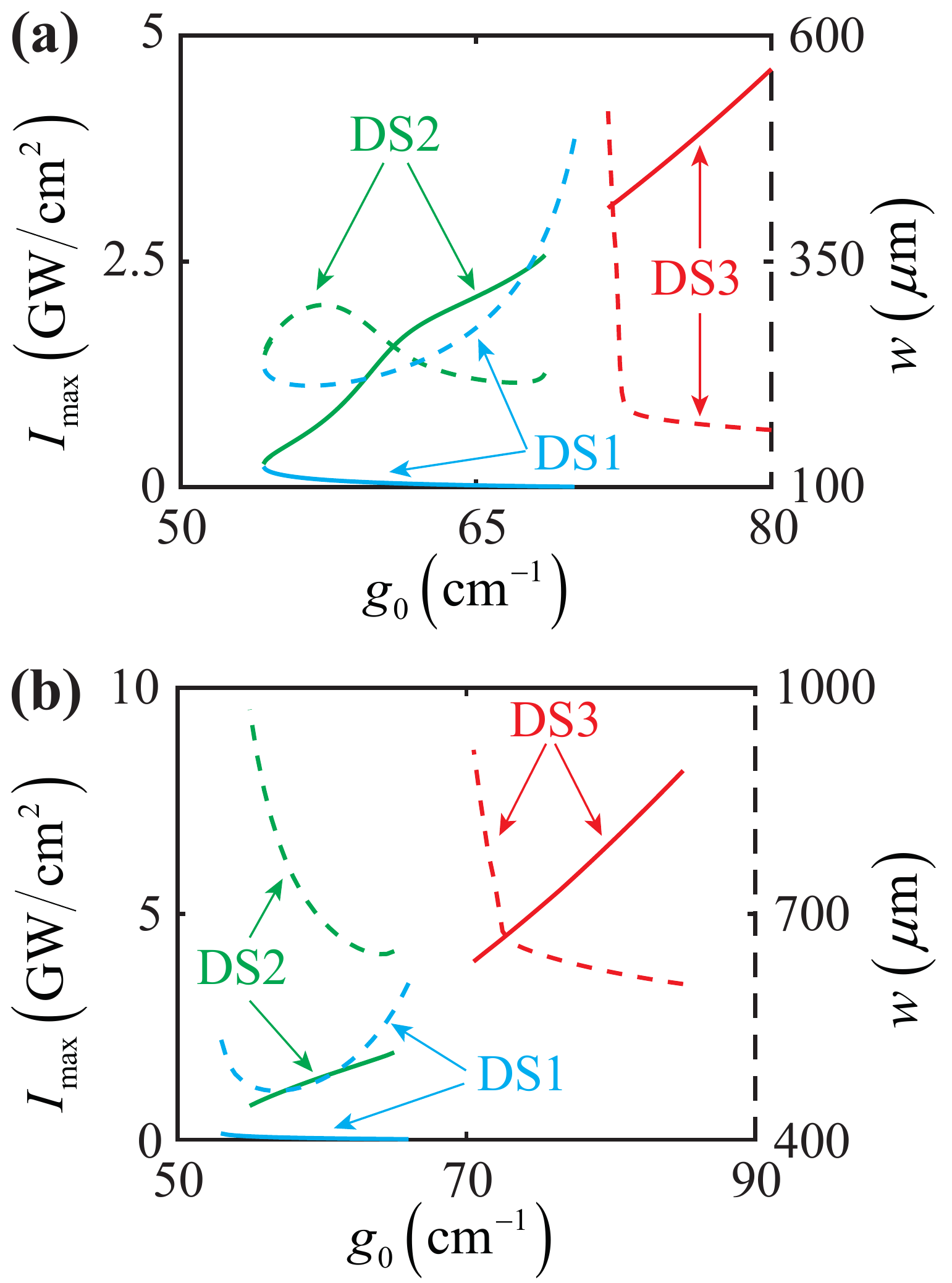}
\caption{Dependence over the linear gain coefficient $g_0$ of peak intensities $I_{\rm max}$ (full lines, left full vertical axis) and spot-sizes $w$ 
(dashed lines, right dashed vertical axis) of the numerically calculated DSs with topological charge {\bf (a)} $m=0$ and {\bf (b)} $m=1$ for graphene 
filling fraction $f = 2.13 \times 10^{-4}$.}
\label{Fig5}
\end{center}
\end{figure}

\begin{eqnarray}
&& - \delta\beta A_0(r_{\bot}) + \frac{1}{2\beta_0} \left[ \frac{1}{r_{\bot}}\frac{d}{dr_{\bot}}\left(r_{\bot} \frac{dA_0}{dr_{\bot}} \right) - \frac{m^2}{r_{\bot}^2}A_0\right] + \nonumber \\
&& + \frac{\beta_0}{2\epsilon_{\rm b}} \left[\epsilon_{\rm eff} \left(|A_0(r_{\bot})|^2\right) - \epsilon_{\rm b} \right] A_0(r_{\bot})  = 0.
\end{eqnarray}
In order to solve the ODE above, we transform it into a nonlinear system of algebraic equations by discretizing the spatial variable $r_{\bot}(n) = r_n$, 
first-order $d A_0 / d r_{\bot} = [A_0(r_{n+1})-A_0(r_n)]/(r_{n+1} - r_n)$ and second-order 
$d^2 A_0 / d r_{\bot}^2 = [A_0(r_{n-1})-2A_0(r_n)+A_0(r_{n+1})]/(r_n - r_{n-1})^2$ derivatives with $n = 1,2,..,N$ and by applying the 
homogeneous boundary conditions $A_0(r_1) = A_0(r_2)$ and $A(r_N) = 0$. We solve the resulting nonlinear system of algebraic equations numerically through the 
Newton-Raphson method. Also for localized NWs, we observe a subcritical bifurcation from the trivial vacuum $A_0({\bf r}_{\bot}) = 0$ and we find three types 
of localized stationary solutions (DS1, DS2, DS3) for every $m=0,1,2$. In Figs. \ref{Fig5}a,b we plot the maximum intensity 
$I_{\rm max} = (1/2) \epsilon_0 \sqrt{\epsilon_{\rm b}} c ~ {\rm max} |A_0(r_{\bot})|^2$ (full lines) and the spot-size 
$w = 2 [\int_{0}^{+\infty} r_{\bot}^3 |A_0(r_{\bot})|^2 d r_{\bot} / \int_{0}^{+\infty} r_{\bot} |A_0(r_{\bot})|^2 d r_{\bot} ]^{1/2}$ (dashed lines)
of (a) $m=0$ and (b) $m=1$ DSs (of types $1,2,3$) against $g_0$ for fixed graphene filling fraction $f = 2.13 \times 10^{-4}$. Similarly to HNWs, also for 
localized NWs DS1 and DS2 coexist in the bistable subcritical domain, while DS3 exists only in the overcritical domain. All the intensity profiles of the DSs 
found are bell-shaped, while their phase $\varphi({\bf r}_{\bot}) = {\rm atan} [{\rm Im} A_0({\bf r}_{\bot}) / {\rm Re} A_0({\bf r}_{\bot})]$ is not uniform 
over the $x-y$ plane, which implies an internal power flow enabling stationary propagation of DSs \cite{Grelu2012}. Indeed, note that in the considered 
dissipative system, conversely to traditional soliton families in Hamiltonian systems, for every DS type there exists only one solution with fixed peak intensity 
and spot-size, which arises from the double nonlinear compensation of diffraction vs focusing and gain vs absorption \cite{Grelu2012}. Note also that both the peak
intensity ($0<I_{\rm max}<10$ GW$/$cm$^2$) and the spot-size ($100$ $\mu$m $<w<10$ $1$ mm) of every DS can be tuned efficiently by the pump intensity $I_{\rm pump}$ modulating the linear gain coefficient $50$ cm$^{-1}<g_0(I_{\rm pump})<90$ cm$^{-1}$, which is experimentally attainable with R6G \cite{Noginov2008}. 

\begin{figure}[t]
\centering
\begin{center}
\includegraphics[width=0.5\textwidth]{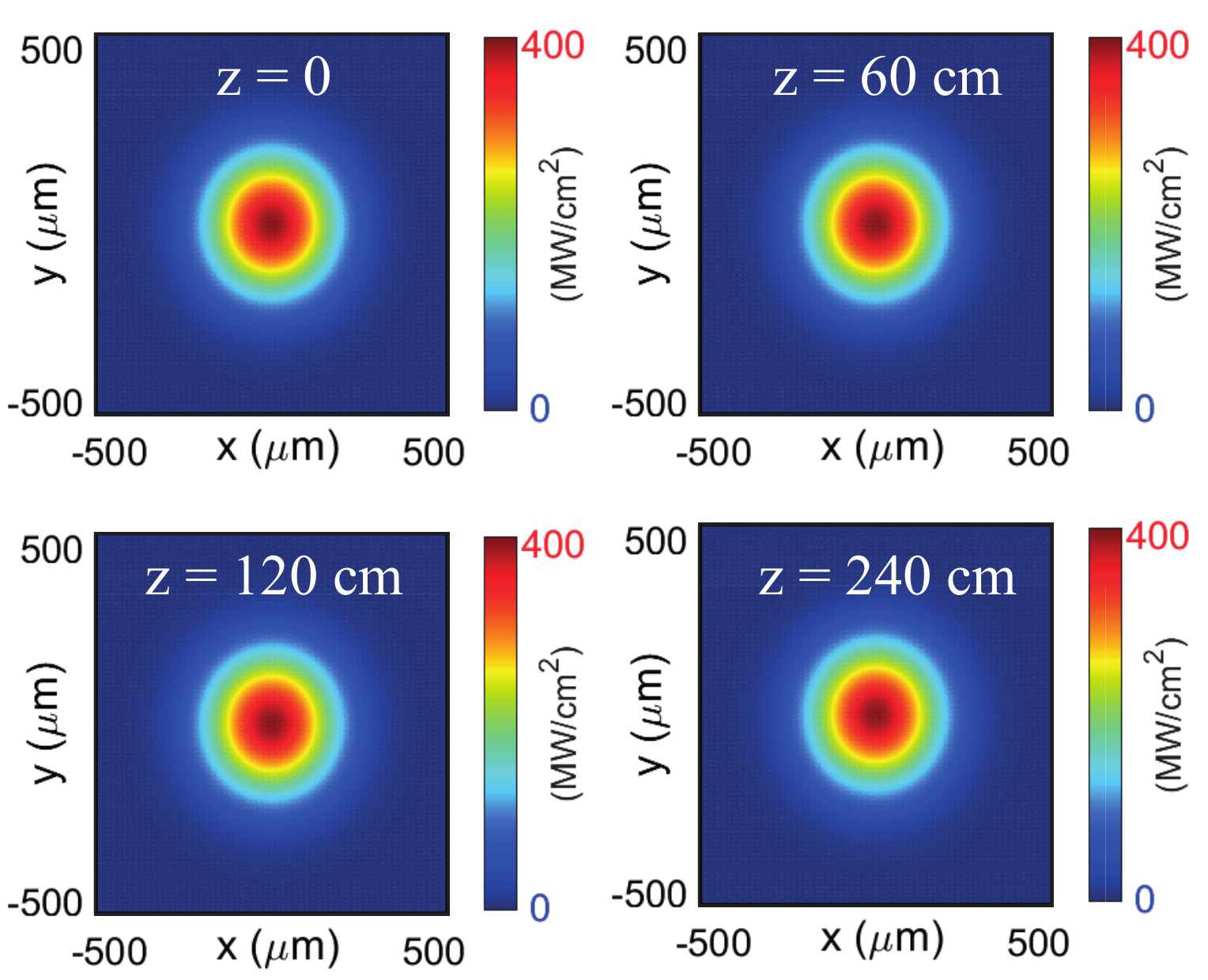}
\caption{Density plots illustrating the dependence of the intensity profiles $I({\bf r}_{\bot},\bar{z})$ of the stable $m=0$ DS2 (for $g_0 = 55$ cm$^{-1}$ 
and $f = 2.13 \times 10^{-4}$) over ${\bf r}_{\bot} = x\hat{e}_x + y\hat{e}_y$ and stationary propagation at several distinct $z=\bar{z}$ indicated on top of 
every plot. }
\label{Fig6}
\end{center}
\end{figure}

\begin{figure*}[t]
\centering
\begin{center}
\includegraphics[width=\textwidth]{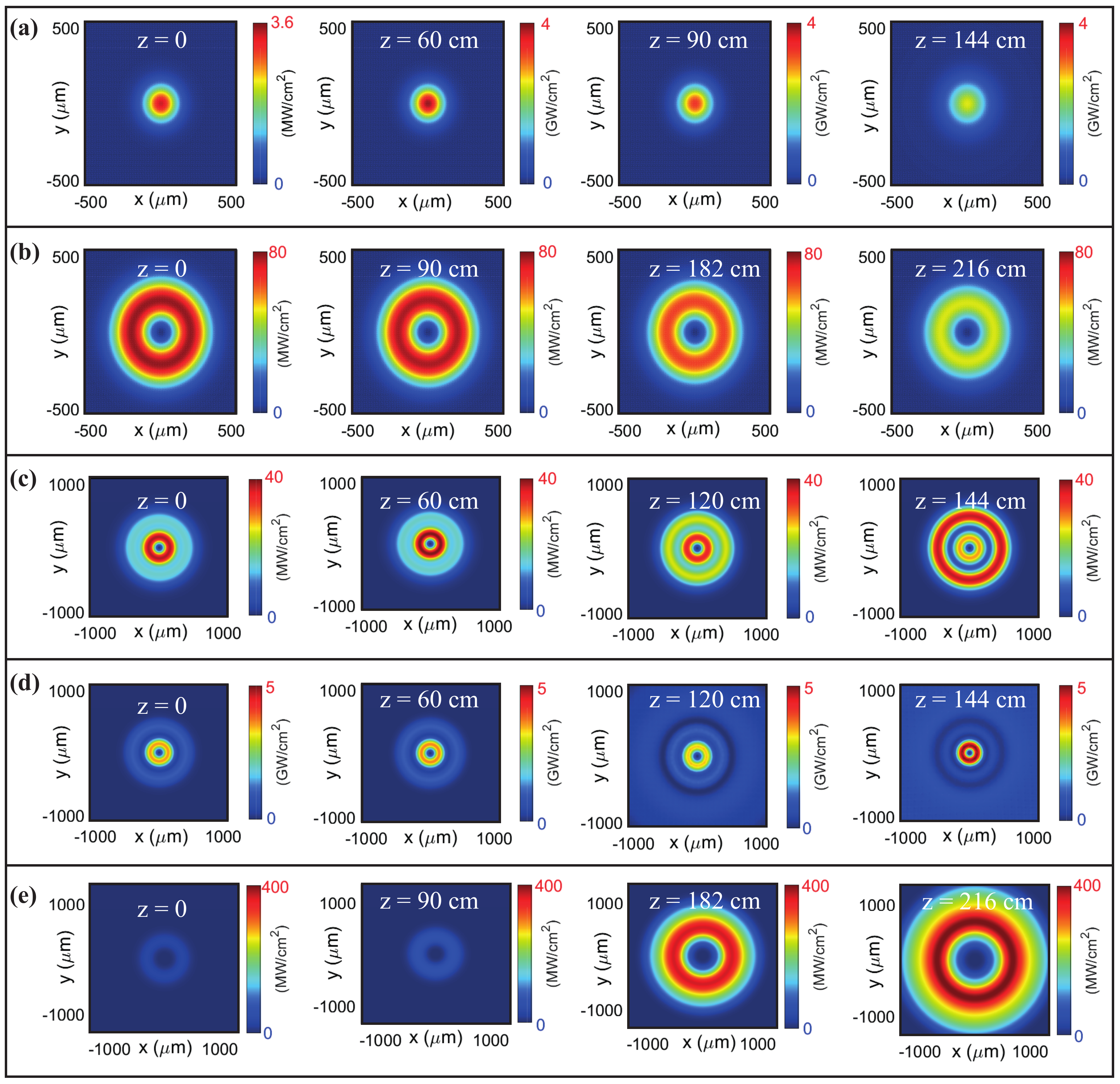}
\caption{Density plots illustrating the dependence of the intensity profiles $I({\bf r}_{\bot},\bar{z})$ over ${\bf r}_{\bot} = x\hat{e}_x + y\hat{e}_y$ 
of (a) $m=0$ DS3 for $g_0 = 75$ cm$^{-1}$, (b) $m=1$ DS1 for $g_0 = 55$ cm$^{-1}$, (c) $m=1$ DS2 for $g_0 = 60$ cm$^{-1}$, (d) $m=1$ DS3 for $g_0 = 75$ cm$^{-1}$, 
and $m=2$ DS2 for $g_0 = 60$ cm$^{-1}$ at several distinct $z=\bar{z}$ indicated on top of every plot. In all plots the graphene filling fraction is set to $f = 2.13 \times 10^{-4}$.}
\label{Fig7}
\end{center}
\end{figure*}

\section{Discussion}

In order to shed light on radiation dynamics in the complex system considered, we investigate the stability of the found DSs for $m=0,1,2$ in propagation by 
solving Eq. (\ref{PropEq}) through a split-step discrete Fourier transform complemented with a fourth-order Runge-Kutta algorithm. We find that stability only 
occurs for $m=0$ DS2 in the subcritical domain, which intensity profile and stationary propagation is illustrated in Fig. \ref{Fig6}, while all the other DSs found
($m=0$ DS1,3 and $m=1,2$ DS1-3) are unstable. In Fig. \ref{Fig7} we illustrate the initial steps of the unstable dynamics of (a) $m=0$ DS3 for $g_0 = 75$ cm$^{-1}$,
(b) $m=1$ DS1 for $g_0 = 55$ cm$^{-1}$, (c) $m=1$ DS2 for $g_0 = 60$ cm$^{-1}$, (d) $m=1$ DS3 for $g_0 = 75$ cm$^{-1}$, and $m=2$ DS2 for $g_0 = 60$ cm$^{-1}$. In all
plots the graphene filling fraction is set to $f = 2.13 \times 10^{-4}$. 
Due to the instability of the background and the existence of several unstable DSs, propagation over longer distances becomes chaotic in the overcritical domain.
Conversely, propagation over longer distances in the subcritical domain leads to filamentation into several stable $m=0$ DS2 solitons, see the supplementary videos, 
where we illustrate the breaking of $m=1$ and $m=2$ DSs into $2$ and $4$ repulsive $m=0$ DSs, respectively.

\section{Conclusions}

In conclusion, we have described the spatial formation of dissipative nonlinear waves in a graphene-based active random metamaterial composed of randomly 
dispersed graphene nano-flakes and optically pumped R6G. Owing to bistability of HNWs in a subcritical region below the instability threshold of the background,
there exist several DS vortices with distinct topological charges. By investigating systematically the existence and stability of DSs one finds that only one
specific DS with vanishing topological charge is stable in the subcritical domain and becomes unstable in the overcritical domain. Analysis of propagation dynamics 
of perturbed HNWs and DSs indicates that nonlinear waves break into stable and repulsive DSs with fixed topological charge $m=0$. This indicates that graphene-based
active random metamaterials enable the control of mode operation since the spatial pattern of the excited nonlinear waves is fixed by the density of graphene 
nano-flakes and the external optical pump tuning the linear gain coefficient $g_0$. Our results indicate that self-organization of coherent radiation structures 
in disordered systems can be exploited to design cavity-free laser operation and and amplification, opening novel possibilities for the advancement of
random lasers.

\end{document}